\documentclass[12pt]{iopart}
\usepackage{graphicx}
\begin{document}

\title{RSA Study of Dimers}

\author{Michal Ciesla$^{1}$ and Jakub Barbasz$^{1,2}$}

\address{$^1$ M. Smoluchowski Institute of Physics, Jagellonian University, 30-059 Kraków, Reymonta 4, Poland.}
\address{$^2$ Institute of Catalysis and Surface Chemistry, Polish Academy of Sciences, 30-239 Kraków, Niezapominajek 8, Poland.}
\ead{michal.ciesla@uj.edu.pl}
\begin{abstract}
The first theoretical study of a dimer’s adsorption process at homogeneous surface is presented. By using the RSA algorithm, we show example monolayers, discuss estimations of random jamming coverage and measure the surface blocking function, which could be used for calculating real systems kinetics. We also found the correlation function for generated coverages and analysed orientational ordering inside the adsorbed monolayer. Results were compared with theoretical and experimental data.
\end{abstract}

\maketitle

\section{Introduction}

Adsorption (in particular irreversible adsorption) of bio-particles at interfaces plays an extremely important role in biotechnology, medicine, chemistry and environmental technology. Some common examples are: paper production \cite{bib1}, particle deposition controlling \cite{bib2} and selective deposition of cells and viruses \cite{bib3, bib4}. Simultaneously, there has been much interest focused on patterned surfaces with regular shape features like circles and dots, squares, stripes and others \cite{bib4, bib5, bib6, bib7}. Those works typically have been based on theoretical adsorption models calculated for convex particles like spheres, spheroids \cite{bib8} and spherocylinders \cite{bib9}. However, the majority of protein molecules are not convex and, therefore, there has been recent interest in the packing of concave objects \cite{bib10, bib11, bib12, bib13, bib14}. It appears that particle shape and symmetry can be crucial for fundamental properties of adsorbed monolayer \cite{bib15, bib16, bib17, bib17b}. In this paper we study the irreversible deposition process of a dimer -- the simplest, concave particle that could model plenty of bio-molecules.

The following section outlines the model, algorithm and parameters used during the simulation of the dimer’s deposition process. The next part contains results and discussion. It is focused mainly on the random maximal coverage ratio. The ratio’s estimation accounts for several theoretical models including blocking function analysis. This section also describes autocorrelations and orientational ordering inside the covering layer. The paper ends with a short summary.

\section{Model and simulation procedure}

\begin{figure}[htb]
\begin{center}
\includegraphics[width=6cm]{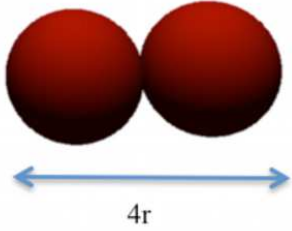}
\end{center}
\caption{Model of a single dimer. In the following considerations we treat r as a~length unit (r=1).}
\label{fig:dimer}
\end{figure}

The adsorption process typically takes place when colloidal particles diffuse close to the surface. Due to adhesion this process can create a film consisting of randomly adsorbed molecules. Here we are focused on irreversible adsorption producing monolayers of adsorbate. The most straightforward approach to numerically simulate these processes is molecular dynamics (MD). The advantages of MD are prediction accuracy and control over most environmental parameters like temperature and the diffusion constant. The main drawback is performance. For this reason we decided to use another method known as continuum Random Sequential Adsorption (RSA), which had been successfully applied to study colloidal systems \cite{bib18}. It is based on independent, repeated attempts to add a dimer to film.
The features of a single step of numerical procedure used here were the following:
\begin{description}
\item[-] a virtual particle was created and its position and orientation on a collector was chosen randomly, accordingly to the uniform probability distribution.
\item[-] an overlapping test with adsorbed earlier nearest neighbours of a virtual particle was performed. This test bases on checking if a surface-to-surface distance between particles is greater than zero.
\item[-] if there was no overlap the virtual particle was adsorbed and added to an existing covering layer. Its position did not change during further calculations, which reflected irreversibility of the process.
\item[-] if there was overlap the virtual particle was removed and abandoned.
\end{description}
Attempts are repeated iteratively. Their number is typically expressed using a~dimensionless time unit:
\begin{equation}
t_0 = N\frac{S_D}{S_C}
\end{equation}
where $N$ is a number of attempts, $S_D = 2\pi r^2$ stands for the coverage given by a single dimer (\fref{fig:dimer}) and $S_C$ is a collector area.
It is worth noticing that there is at least one RSA algorithm allowing faster saturation of the underlying collector \cite{bib:Torquato2006}. It bases on tracing uncovered area and uses this information to decrease overlapping probability for forthcoming RSA attempts. However we did not decide to use it here, because it complicates analysis of the standard RSA kinetics, which we want to compare with previous works.

\begin{figure}[htb]
\begin{center}
\includegraphics[width=14cm]{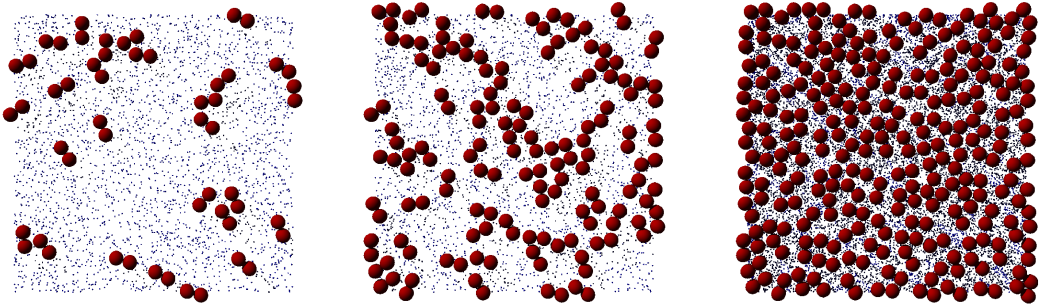}
\end{center}
\caption{Typical monolayer samples for three different coverages: $\theta=0.1$, $\theta=0.3$ and $\theta \approx 0.55$. The collector side length was equal to $40 r$. Fixed boundary conditions were used.}
\label{fig:layer}
\end{figure}

\begin{figure}[htb]
\vspace{20pt}
\begin{center}
\includegraphics[width=8cm]{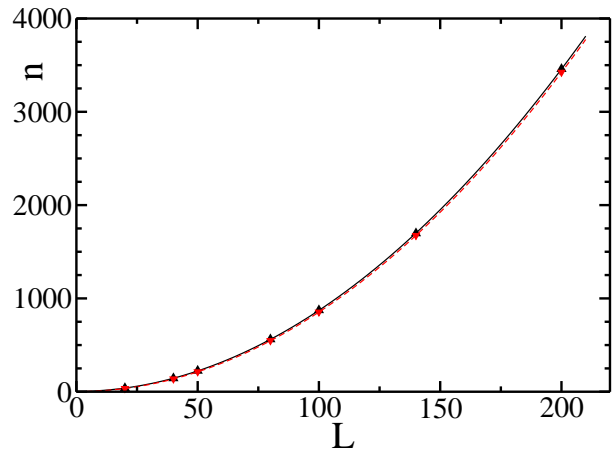}
\end{center}
\caption{Number of adsorbed dimers versus collector side length. Signs represent values taken from simulations for fixed boundaries (triangles up) as well as for periodic boundary conditions (triangles down). Lines are quadratic fits: $n(L)= 0.086 L^2+0.115 L+2.59$ (solid line for fixed boundaries) and $n(L)= 0.086 L^2-0.066 L+1.654$ (dashed line for periodic boundary conditions).}
\label{fig:nsize}
\end{figure}

In the case of our simulations, the adsorption was stopped after $T=10^5 t_0$. Collector sizes used varied from 20 to 200 $r$. Simulations were performed using fixed boundary collectors as well as collectors with periodic boundary conditions. In the first case centres of both circles forming a dimer had to be inside a collector area. For each collector we get at least $100$ covering layers. The coverage ratio $\theta$ is calculated as follows:
\begin{equation}
\theta = n_d \frac{S_D}{S_C},
\end{equation}
where $n_d$ is a number of adsorbed dimers. Typical coverages for three different values of coverage ratio $\theta$: are presented in \fref{fig:layer}.
Looking at the above drawing for jamming coverage ($\theta \approx 0.55$), one can notice that the dimers density seems to be higher near the collector edges. This is an undesirable effect as we are mainly interested in the coverage ratio for an infinite collector. In order to control the systematic error caused by finite size and boundary conditions, the whole process was simulated over different sized collectors with fixed boundaries as well as with periodic boundary conditions. Results shown in \fref{fig:nsize} suggest that the bias of a measure (deviation from a pure quadratic fit) is less than 1\% for the largest collector we used, independently on the specific boundary conditions used. On the other hand, for small collectors with fixed boundaries, higher densities near collector edges can be successfully used for producing systems with interesting optical properties such as micro-lenses.

Most of results discussed later in the paper were obtained using largest collector ($L=200$) with fixed boundaries. We checked that use of periodic boundary conditions does not have any significant influence on presented conclusions.

\section{Results and discussion} 

\subsection{The maximal random coverage ratio}
\label{sec:coverage}

\begin{figure}[htb]
\vspace{20pt}
\begin{center}
\includegraphics[width=8cm]{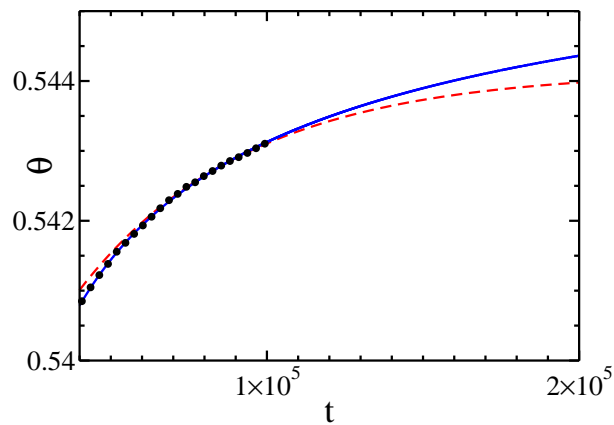}
\end{center}
\caption{Coverage versus dimensionless time (\ref{eq3}). Dots come from simulation. Lines represent example kinetisc fits, solid: $\theta_{max} - \theta(t) \sim t^{-1/2}$ and dashed: $\theta_{max} - \theta(t) \sim \exp(-ct)$.}
\label{fig:coverage}
\end{figure}

\begin{figure}[htb]
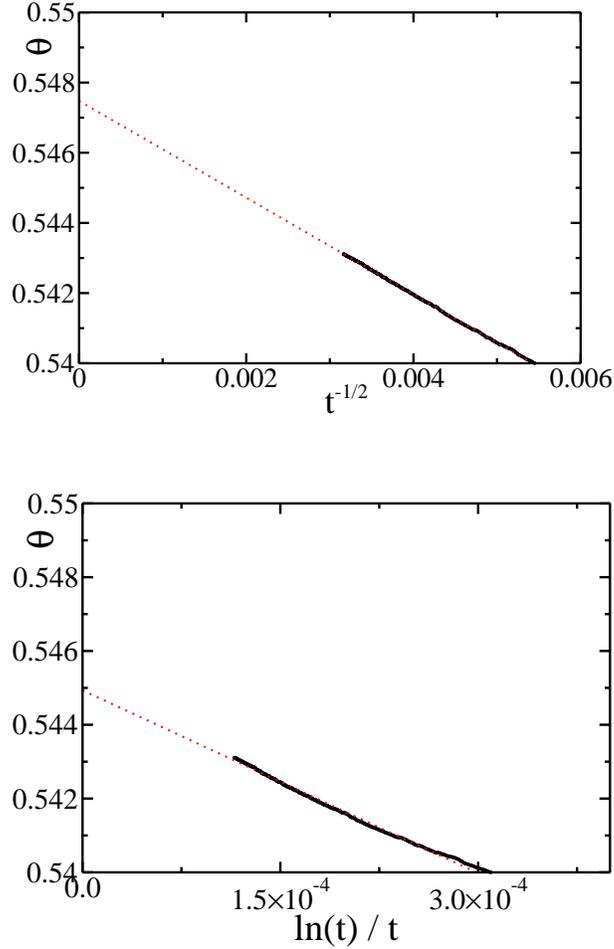

\vspace{20pt}
\begin{center}
\includegraphics[width=8cm]{t_-0.5.eps}
\begin{verbatim}
\end{verbatim}
\includegraphics[width=8cm]{lnt_t.eps}
\end{center}
\caption{Coverage versus $t^{-1/2}$ and $\ln t / t$. Bold points are taken from simulation. Dotted lines represent linear fits: $\theta_{max}-\theta = 0.54748 - 1.3787\cdot t^{-1/2}$ and $\theta_{max}-\theta = 0.54493 - 16.512\cdot \ln t / t$ respectively.}
\label{fig:comparison}
\end{figure}

The main purpose of this work was to determine the maximal random adsorption ratio for dimers and compare it with results obtained for hard circles (spheres) \cite{bib16, bib:Torquato2006,  bib:Swendsen1981, bib:Privman1991}. That ratio should be specified for an infinite collector area and infinite adsorption time. Despite controlling the error due to finite collector size one have to deal with finite simulation times. Particularly, in the case of large collectors, it is not certain if there is any possibility of adsorption after the simulation time and therefore approximation of maximal coverage depends on the RSA kinetics model (see fig.\ref{fig:coverage}).
There were plenty of previous works in this area \cite{bib18, bib:Swendsen1981, bib:Privman1991, bib19,bib20,bib21,bib22} and the general conclusion is that asymptotically:
\begin{equation}
\label{eq:t12}
\theta_{max} - \theta(t) \sim t^{-1/D}.
\end{equation}
for irreversible deposition of circles or unoriented squares. $D$ here denotes collector dimension. The situation changes when adsorbed particles are ordered \cite{bib:Swendsen1981, bib:Privman1991}. For example deposition of oriented squares for long enough time obeys the following relation:
\begin{equation}
\label{eq:lntt}
\theta_{max} - \theta(t) \sim \frac{(\ln t)^{D-1}}{t}.
\end{equation}
In case of present work the planar orientation of particles were chosen randomly with an uniform probability distribution. However this symmetry could be broken because in close proximity of previously adsorbed particle there are more space for parallely aligned particles than for perpendicular ones. Therefore asymptotic (\ref{eq:t12}) and (\ref{eq:lntt}) were compared in fig.\ref{fig:comparison}.
Although both relations approximates the experimental data well, the fit (\ref{eq:t12}) is slightly better in terms of linear correlation coefficient. The values of $\theta_{max}$ can be obtained by interpolation of $t^{-1/2}$ and $\ln t / t$ to $0$. Here they are $\theta_{max}=0.5475$ and $\theta_{max}=0.5449$ respectively. The difference is located within a 1\% systematic margin of error coming from finite collector size.

In order to compare our result with experiments, we analysed data obtained for adsorption of insulin. Assuming insulin dimer has a mass of $11616 [Da]$ and a size of $875 [\AA^2]$ one can find that $\theta_{max}=0.55$ corresponds to a monolayer surface density of $1.21 [mg/m^2]$. Typical values from experiments are in the range of $1.3-1.6 [mg/m^2]$ depending on concentration and insulin type (human, Zn-Free) \cite{bib23,bib24}. Difference is noticeable but not significant. It can be explained by two causes: a more regular shape of the insulin dimer compared to our model (\fref{fig:dimer}) and the fact that, depending on concentration, the insulin particle can also appears as a monomer or hexamer. Both of them follow to higher coverages.

\subsection{Blocking function and fluctuations in the number of adsorbed particles}

In the real experiment, adsorption kinetics depends typically on two factors: efficiency of the transport process (mainly diffusion or convection – depending on experimental setup) that brings adsorbate from the bulk to the surface and the probability of catching particles, which are in a close proximity \cite{bib8,bib19,bib21,bib22,bib25,bib26,bib27,bib28,bib29}. Here we would like to focus on the second factor, which is described by the blocking function, also known as the Available Surface Function (ASF). It can be easily obtained from the simulation as a ratio of successful attempts to all RSA attempts. 
\begin{figure}[htb]
\vspace{20pt}
\begin{center}
\includegraphics[width=8cm]{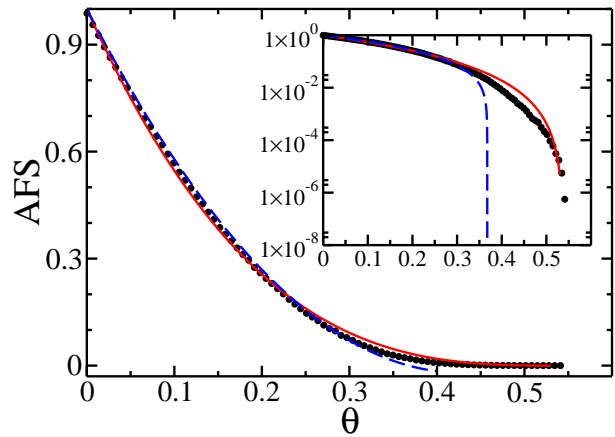}
\end{center}
\caption{The blocking function - successful attempts ratio versus coverage. Dots are simulation data, whilst solid and dashed lines are fits: $ASF(\theta )=1-4.77 \theta + 5.58  \theta^2$, and $ASF(\theta )=(1-\theta /0.543)^{2.97}$ correspondingly. The inset presents the same data in a logarithmic scale.}
\label{fig:blocking}
\end{figure}
Obtained ASF is presented in \fref{fig:blocking}. The quadratic fit is commonly used to estimate adsorption rates at a limit of small coverages:
\begin{equation}
\label{eq:asf}
ASF(\theta) = 1 - C_1 \theta + C_2 \theta^2
\end{equation}
In the case of dimers, simulations show that $C_1=4.77$ and $C_2=5.58$, whereas those parameters for hard circles adsorption are $C_1=4$ and $C_2=3.31$ \cite{bib21}. It shows that available surface shrinks faster for dimers. This follows intuition, because for successful adsorption a particle should have not only appropriate coordinates but also the right orientation. 

Moreover, ASF provides another way to estimate maximal coverage by analysing an adsorption probability. Specifically, the second fit in \fref{fig:blocking} suggests that maximal coverage is equal to $\theta_{max}=0.543$ and also provide additional support for model (\ref{eq:t12}) because:
\begin{equation}
\frac{d \theta(t))}{d t} = ASF(\theta(t)) = \left( 1 - \frac{\theta(t)}{\theta_{max}} \right)^3
\end{equation}
yields
\begin{equation}
\theta_{max} - \theta (t) \sim t^{-1/2}
\end{equation}

\begin{figure}[htb]
\vspace{20pt}
\begin{center}
\includegraphics[width=8cm]{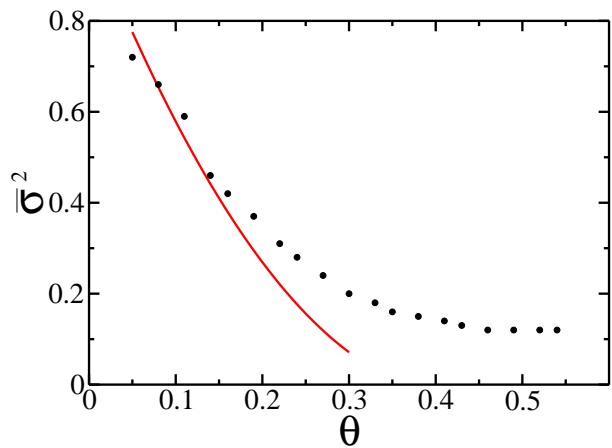}
\end{center}
\caption{Reduced variance of adsorbed particles as a function of a coverage. Dots correspond to the simulation data, whilst solid line is a quadratic fit (\ref{eq:asf}) to ASF.}
\label{fig:sigma}
\end{figure}

Although estimating adsorption kinetics through ASF is straightforward and commonly used, there is a problem with direct measure of the blocking function during experimentation. Therefore, researchers are likely to count density fluctuations in the number of adsorbed particles for a given coverage. Typically, those fluctuations are expressed in term of reduced variance of particle number $n$ inside a specified area: $\bar{\sigma}^2 = \sigma^2 (n) / \left< n \right>$. It can be shown that, at least in the limit of small coverages, $\bar{\sigma}^2(\theta )=ASF(\theta )$ \cite{bib30}. This comparison, in the case of our simulations, is presented in \fref{fig:sigma}. 
As expected, $\bar{\sigma}^2(\theta )$ follows $ASF(\theta )$ only for small coverages. For $\theta > 0.2$ the difference between them becomes significant.

\subsection{Autocorrelations}

\begin{figure}[htb]
\vspace{20pt}
\begin{center}
\includegraphics[width=8cm]{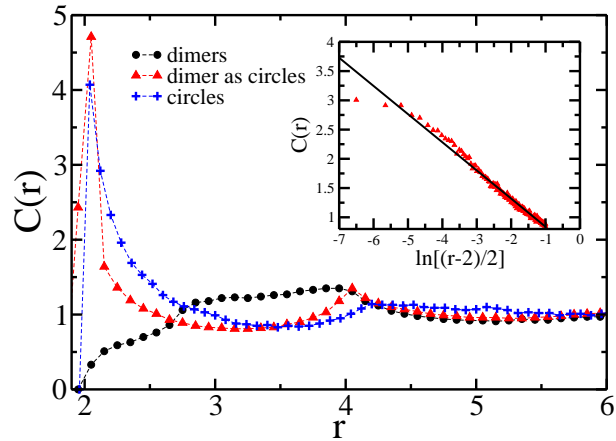}
\end{center}
\caption{Autocorrelation functions. The data were analysed in two different ways. Dots represents results for dimers as themselves whereas triangles shows autocorrelation for circles, regardless of that they are forming a dimer or not. Pluses show the reference results for circles. Inset shows asymptotic behaviour for small $r$. Fitted line is $C(r) = - 0.48151 \cdot \ln[(r-2)/2] + 0.35683$.}
\label{fig:cor}
\end{figure}
Autocorrelation of particles is another important characteristic of the monolayer. Here we are able to measure at least two different autocorrelation. The first is a standard distribution of distance between two molecules. Results are shown in \fref{fig:cor}. Let’s first concentrate on reference frame given by autocorrelation inside monolayer build of spherical particles (pluses in \fref{fig:cor}). The function has a maximum for $r=2.0$ (nearest possible neighbour), then due to excluded volume it approaches minimum. Next, a much weaker maximum is around $r=4$. Because of the random character of coverage, those oscillations vanish superexponentially \cite{bib:Torquato2006} and functions stabilises at a value of $1$, which is a result of normalisation.

Autocorrelation for dimers (circles) is different. Function rises very slowly with distance to approach its first maximum around $r=3.8$. It suggests for example that there are not many dimers lying side-by-side, at a distance close to $r \approx 2$. Then, correlation approaches wide and shallow minimum around $r=5$.  This behaviour is mainly an effect of a dimer’s shape. We expect that for more anisotropic molecules, for example fibrinogens, there could be no minimum at all. 

Although at first it seems that autocorrelations for circles and dimers are totally different they can be easily compared when dimers coverage is treated as it was build up of independent circles. Here (triangles) autocorrelation looks almost the same as in the case for circles, only that maxima and minima are sharper. This similarity could explain why maximal random coverage for circles and dimers are so close. The asymptotic behaviour is also similar.  At small distances it has the universal form derived in \cite{bib:Swendsen1981, bib:Privman1991}:
\begin{equation}
\label{eq:cor}
C(r) \sim -\ln[(r/2-1)] \,\,\,\, \mbox{for} \,\,\, r \to 2^+ ,
\end{equation}
whereas for large $r$ the decay seems to be even faster than for circles.

\subsection{Ordering}

The non-uniform shape of a dimer gives the possibility to check if any orientational order appears in a monolayer. Such ordering was widely investigated before, but mainly using lattice topology of collector surface eg. \cite{bib31}. As mentioned in sec. \ref{sec:coverage} it could also influence on kinetics of RSA.
 
\begin{figure}[htb]
\vspace{20pt}
\begin{center}
\includegraphics[width=8cm]{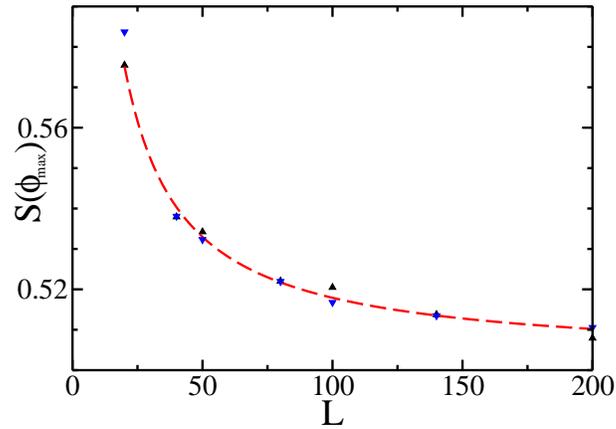}
\end{center}
\caption{Order $S(\phi_{max} )$ dependence on collector side size. Trangles represent simulation data for fixed boundaries (triangles up) and periodic boundary conditions (triangles down). Dashed line is a simple analytical fit: $S(\phi_{max} )=1.58/L+2.45/L^2 +0.5$.}
\label{fig:ordersize}
\end{figure}

\begin{figure}[htb]
\vspace{20pt}
\begin{center}
\includegraphics[width=8cm]{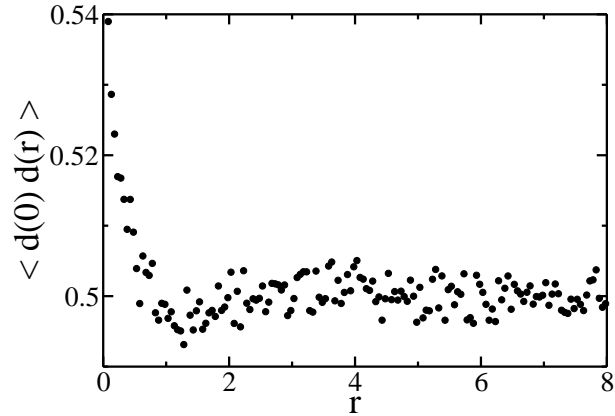}
\end{center}
\caption{Correlation of ordering  versus distance for collector size $200 x 200$.}
\label{fig:order}
\end{figure}

To measure orientational order in our, continuous system we introduce the following function determined by a dimer’s configuration:
\begin{equation}
S(\phi) = \frac{1}{N} \sum_{i=1}^N \left( x_i \cos \phi + y_i \sin \phi \right)^2,
\end{equation}
where $(x_i,y_i )$ are coordinates of a unit vector along the $i$-th molecule in a layer. It can be noticed that $S(\phi)$ is an average scalar product between molecules orientation and the direction given by an angle $\phi$. Thus, for an ideally aligned layer, $S(\phi)$ will oscillate between $0$ and $1$, where the maximum corresponds to an angle being parallel to molecules and minimums are reached for $\phi$ perpendicular to the direction of alignment. For pure random alignment $S(\phi)$ will be constant and equal to $0.5$. In general, the mean orientation given by maximum $S(\phi)$ can be estimated from:
\begin{equation}
\tan \phi_{ex} = \frac{\sum_{i=1}^N x_i \cdot y_i}{\sum_{i=1}^N x_i \cdot x_i - \sum_{i=1}^N y_i \cdot y_i}.
\end{equation}
The above equation is fulfilled by both maximum $\phi_{max}$ and minimum $\phi_{min}=\phi_{max} + \pi/2$. Dependence between maximal value of $S(\phi )$ and collector size is shown in \fref{fig:ordersize}.
Global order, although not very high, are strongest in small collector coverages. There could be at least two reasons for this. Firstly, if the allowed space for placing the following dimer is restricted, it is more probable to find enough room for parallel alignment than for a perpendicular one. Secondly, our adsorption conditions prefer parallel alignment at collector borders, because the dimer is placed down only when centres of two of their circles are touching the surface. For small collectors both, described above, effects are stronger however influence of fixed boundaries is irrelevant (fig.\ref{fig:ordersize}).  In order to determine which one of them is more important, we analysed local ordering in large collectors. It was done by calculating the mean value of scalar product between two dimers at a given distance. Results are drawn in \fref{fig:order}.
The local ordering and its range are quite small. It practically disappears when distance exceeds $0.5$. It suggests that ordering visible in tiny collectors is mainly due to their small size.

\subsection{RSA of dimers in higher dimensions}
The RSA of (hiper) spheres and shows some universal behaviour regardless of collector dimension \cite{bib16, bib:Torquato2006, bib:Swendsen1981, bib:Privman1991}. The best known of them are Feder's law (\ref{eq:t12}) and asymptotic relation for autocorrelation function (\ref{eq:cor}). Results described in previous sections shows that those relations are also valid for dimers in 2D. Moreover the maximal random coverage ratio agrees with value obtained for circles. It is interesting if this is only accidental coincidence or more general feature. To address this question we looked at RSA of dimers in 3D. As a collector we used cube having side length $35$ with a periodic boundary conditions and dimer was modelled by two touching spheres. Presented results are obtained from $50$ independent simulation runs.  

\begin{figure}[htb]
\vspace{20pt}
\begin{center}
\includegraphics[width=8cm]{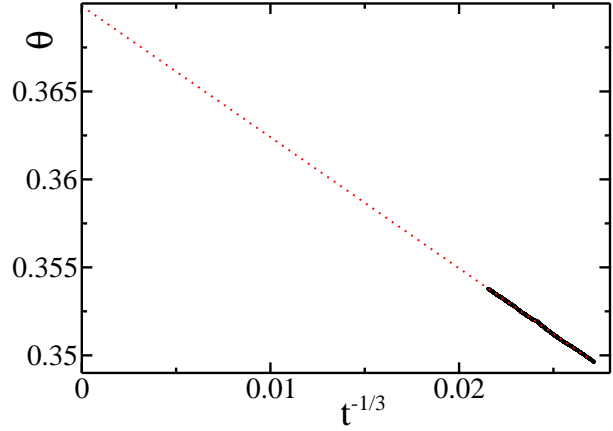}
\end{center}
\caption{Coverage versus $t^{-1/3}$. Bold points are taken from simulation. Dotted lines represent linear fits: $\theta_{max}-\theta = 0.36986 - 0.74547\cdot t^{-1/3}$.}
\label{fig:federslaw3d}
\end{figure}

Figure \ref{fig:federslaw3d} presents coverage kinetics versus $t^{-1/3}$.
The kinetics (\ref{eq:t12}) fits well to the data. The random coverage for dimers in 3D is $0.37$ and within margin of error agrees with value $0.0381$ obtained earlier for spheres  \cite{bib:Torquato2006}.

\begin{figure}[htb]
\vspace{20pt}
\begin{center}
\includegraphics[width=8cm]{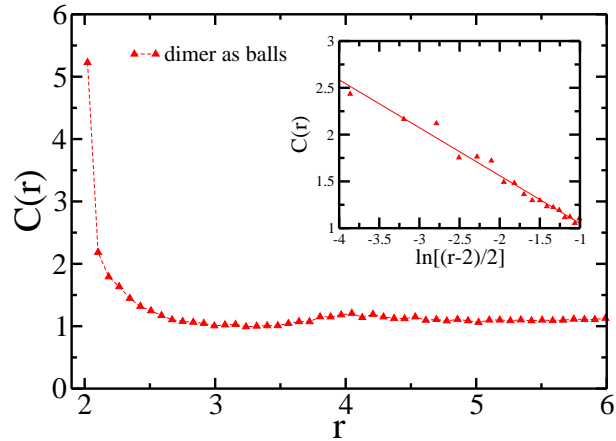}
\end{center}
\caption{Autocorrelation function for spheres, regardless of that they are forming a dimer or not. Inset shows asymptotic behaviour for small $r$. Fitted line is $C(r) = - 0.51176 \cdot \ln[(r-2)/2] + 0.5384$.}
\label{fig:cor3d}
\end{figure}

Autocorrelation function is presented in fig.\ref{fig:cor3d}
Again we observe the same behaviour as in 2D and as for spheres.  The decay for larger distances seems to be superexponential. On the other side, for small $r$ the asymptotic agrees with (\ref{eq:cor}). The linear coefficient are approximately two times smaller than values published for spheres random adsorption both in 2D and 3D \cite{bib:Torquato2006}.

\section{Conclusions}
The maximal random coverage for a dimer’s monolayer is $0.547$ and within the margin of error is not distinguishable from results obtained previously for adsorption of spherical particles. Also monolayer density autocorrelations are almost the same. On the other hand, calculated surface blocking function (ASF) is significantly different ($C_1$ and $C_2$ coefficient). It suggests other kinetics of adsorption, however the Feder's law (\ref{eq:t12}) is maintained. Density fluctuations can successfully estimate ASF only for small coverages. Orientational ordering is imperceptible for macroscopic collectors but could play significant role in micro scale.

Presented results for both maximal random coverages and autocorrelation function suggests that random packing problems for dimers and for spheres is governed by the same rules for $D \ge 2$.

This work was supported by grant MNiSW/0013/H03/2010/70.

\end{document}